\documentclass[a4paper,aps,twocolumn,nofootinbib]{revtex4}
\RequirePackage[colorlinks,hyperindex]{hyperref}
\RequirePackage[english]{babel}
\RequirePackage[latin1]{inputenc}
\RequirePackage[T1]{fontenc}
\RequirePackage{mathrsfs}
\RequirePackage{amsmath}
\RequirePackage{amssymb}
\RequirePackage{amsbsy}
\RequirePackage{color}
\RequirePackage{bm}
\hypersetup{colorlinks=true,breaklinks=true,urlcolor=blue,linkcolor=red}
\begin{document}
\title{\bf{Geometry, Zitterbewegung, Quantization}}
\author{Luca Fabbri}
\affiliation{DIME, Sez. Metodi e Modelli Matematici, Universit\`{a} di Genova, 
Via all'Opera Pia 15, 16145 Genova, ITALY}
\date{\today}
\begin{abstract}
In the most general geometric background, we study Dirac spinor fields with particular emphasis given to the explicit form of their gauge momentum and the way in which this can be inverted so to give the expression of the corresponding velocity; we study how zitterbewegung affects the motion of particles, focusing on the internal dynamics involving the chiral parts; we discuss the connections to field quantization, sketching in what way anomalous terms may be gotten eventually.
\end{abstract}
\maketitle
\section{Introduction}
After nearly seventy years since the first experimental confirmation, quantum field theory (QFT) has yet to fail a phenomenological test. Whether it is the correction to the magnetic moment of fermions or the energy splitting in hydrogen-like systems, QFT has always provided very precise predictions matching to less than a part in several trillions the most accurate of measurements.

Nonetheless, despite its success when compared to our observations, QFT is still lacking a proper mathematical definition: conceptual problems start from the fact that all calculations are done by expanding fields in terms of plane waves, which are not square integrable (and in fact they do not really exist); they continue with the fact that in such an expansion the coefficients are reinterpreted as a pair of creation/annihilation operators, which still lack a definition, and for that matter their set of commutation relations might make no sense \cite{sw}; and they end with the fact that for all these calculations we employ the so-called interaction picture, which has been demonstrated not to exist in general in the context of a Lorentz-covariant field theory at all \cite{h}. In the face of these issues, the fact that perturbative expansions do not converge, or that each of their terms is finite only up to a certain regularization or renormalization, looks like a minor problem indeed.

Of all these conceptual issues, the lack of some proper mathematical definition of the creation/annihilation operators was felt particularly by Schwinger, who took this unsatisfactory situation to prompt himself into finding a different formulation for field quantization: his efforts led him to the construction of the so-called source theory \cite{s}.

Nevertheless, again, this is not a solution: Schwinger's source theory is in fact a prototypical version of the path integral, whose measure has never been defined too.

More in general, if we were to go back to the roots of quantization, we would see that the first problem would be involving the use of plane waves, which are not square integrable and as such they can not represent a particle.

This point, however, may constitute a possible avenue for our way out. As it is, QFT might be too dramatic in its first assumption that only plane waves should be used, thus leaving out too much information, so that some of the lost information must be reintroduced, and this could be done through quantization. In other words, quantization fills the gaps left by too strong approximations enforced by having the particle described with plane waves.

Were this the case, any theory of fields employing only the particular solution given by the plane wave but later quantized through some subsidiary conditions should be replaceable by a theory of fields employing general solutions with no subsidiary condition to be implemented.

We do not know whether this is the case. On the other hand, some literature does exist which follows this path: a first example is the one given by Koba and Welton, who faced the treatment of the anomalous magnetic moment of the electron and the Lamb shift, respectively, in terms of semi-classical considerations \cite{k,w}; more systematic is the work of Barut and co-workers, Dowling above all, who study the electron in electrodynamic self-interaction, and with no references to quantization, recovering the above mentioned results, on anomalous magnetic moment and Lamb shift, in general \cite{bd1,bd2}; the deepening on Lamb shift and new results on spontaneous emission/absorption are also presented in \cite{Barut:1992cs, Barut:1983um, Barut:1986ih, Barut:1987hx}; a result on vacuum polarization is also given in \cite{Acikgoz:1993mu} and discussed with the same spirit.

In fact, the concept of vacuum polarization might come helpful in visualization. Because a plane wave describes a freely propagating point particle and quantization accounts for radiative processes involving virtual loops then much in the same way in which quantization fills the gaps left by working with plane waves all virtual loops fill the gaps left by working with point particles; in addition, the surrounding cloud of virtual loops gives an effective size to what would otherwise be a mere point particle.

In the perspective outlined above, any theory of point particles whose dynamics is corrected in terms of virtual loops should be replaceable by a theory of extended fields whose dynamics is comprehensive enough to contain the corrections attributed to the virtual loops.

Although the lack of any exact solution makes it difficult to know what are all the dynamical effects that could replace the corrections due to virtual processes, one of the possibilities that has been considered is zitterbewegung, as addressed by Hestenes \cite{h5}, and with more details by Recami and Salesi \cite{sr1,sr2}. Zitterbewegung is a dynamic effect due to the relative motion between left-handed and right-handed semi-spinor projections of spinors.

As such it cannot be present for a point particle, whose lack of size means lack of internal structures. Still, it can be, and in fact it is, present in general for extended fields, and therefore it does make sense to consider it as what might give rise to dynamical effects, among which some can be mimicking the corrections of virtual loops.

Consequently, having in mind the idea of reproducing quantum corrections, and following the hint that such an endeavour might be done in terms of zitterbewegung, it is wise to start from the most comprehensive dynamics.

In this paper we will do this, defining the most general dynamics for matter fields, deriving some of the possible consequences of zitterbewegung, and seeing what connection there can be with known quantum corrections.
\section{Fundamental Theoretical Generalities}

\subsection{Dirac Spinorial Field}
To begin, we recall that $\boldsymbol{\gamma}^{a}$ are Clifford matrices, from which $\left[\boldsymbol{\gamma}_{a}\!,\!\boldsymbol{\gamma}_{b}\right]\!=\!4\boldsymbol{\sigma}_{ab}$ and $2i\boldsymbol{\sigma}_{ab}\!=\!\varepsilon_{abcd}\boldsymbol{\pi}\boldsymbol{\sigma}^{cd}$ are the definitions of the $\boldsymbol{\sigma}_{ab}$ and the $\boldsymbol{\pi}$ matrix (this matrix is usually indicated as gamma with an index five, but since in the space-time this index has no meaning we use a notation with no index so to avoid confusion): given $\psi$ as a Dirac spinor field, we define the bi-linear quantities given by
\begin{eqnarray}
&M_{ab}\!=\!2i\overline{\psi}\boldsymbol{\sigma}_{ab}\psi
\end{eqnarray}
with
\begin{eqnarray}
&S^{a}\!=\!\overline{\psi}\boldsymbol{\gamma}^{a}\boldsymbol{\pi}\psi\\
&U^{a}\!=\!\overline{\psi}\boldsymbol{\gamma}^{a}\psi
\end{eqnarray}
as well as
\begin{eqnarray}
&\Theta\!=\!i\overline{\psi}\boldsymbol{\pi}\psi\\
&\Phi\!=\!\overline{\psi}\psi
\end{eqnarray}
and which, despite being written only with spinor fields, are all real tensors. From the metric we define the symmetric connection as usual with $\Lambda^{\sigma}_{\alpha\nu}$ and with it we define the spin connection $\Omega^{a}_{\phantom{a}b\pi}\!=\!\xi^{\nu}_{b}\xi^{a}_{\sigma}(\Lambda^{\sigma}_{\nu\pi}\!-\!\xi^{\sigma}_{i}\partial_{\pi}\xi_{\nu}^{i})$ in such a way that with the gauge potential $qA_{\mu}$ we can define
\begin{eqnarray}
&\boldsymbol{\Omega}_{\mu}
=\frac{1}{2}\Omega^{ab}_{\phantom{ab}\mu}\boldsymbol{\sigma}_{ab}
\!+\!iqA_{\mu}\boldsymbol{\mathbb{I}}\label{spinorialconnection}
\end{eqnarray}
needed to write the spinorial covariant derivative 
\begin{eqnarray}
&\boldsymbol{\nabla}_{\mu}\psi\!=\!\partial_{\mu}\psi
\!+\!\boldsymbol{\Omega}_{\mu}\psi\label{spincovder}
\end{eqnarray}
in which for the moment no torsion is defined: the commutator of spinorial covariant derivatives can be used to justify the definitions of space-time and gauge curvature
\begin{eqnarray}
&R^{i}_{\phantom{i}j\mu\nu}\!=\!\partial_{\mu}\Omega^{i}_{\phantom{i}j\nu}
\!-\!\partial_{\nu}\Omega^{i}_{\phantom{i}j\mu}
\!+\!\Omega^{i}_{\phantom{i}k\mu}\Omega^{k}_{\phantom{k}j\nu}
\!-\!\Omega^{i}_{\phantom{i}k\nu}\Omega^{k}_{\phantom{k}j\mu}\\
&F_{\mu\nu}\!=\!\partial_{\mu}A_{\nu}\!-\!\partial_{\nu}A_{\mu}
\end{eqnarray}
which are again in the torsionless case. The Lagrangian we will consider is given according to the standard
\begin{eqnarray}
\nonumber
&\mathscr{L}\!=\!\frac{1}{4}(\partial W)^{2}\!-\!\frac{1}{2}M^{2}W^{2}
\!+\!R\!+\!\frac{1}{4}F^{2}-\\
&-i\overline{\psi}\boldsymbol{\gamma}^{\mu}\boldsymbol{\nabla}_{\mu}\psi
\!+\!XS^{\mu}W_{\mu}\!+\!m\Phi
\label{l}
\end{eqnarray}
with $R$ trace of the space-time curvature and $F^{2}$ square of the gauge curvature and where the generality we temporarily lost when we defined torsionless connections can now be restored by including torsion as an axial vector $W_{\mu}$ with curl given by $(\partial W)_{\mu\nu}$ for the sake of simplicity.

As it has been discussed first of all by Wigner and more recently by Lounesto and Cavalcanti \cite{L,Cavalcanti:2014wia}, spinor fields can be classified in two large classes, in terms of which a spinor such that $\Theta\!=\!\Phi\!=\!0$ is called \emph{singular} and it is the subject of many studies \cite{daSilva:2012wp, Ablamowicz:2014rpa, daRocha:2016bil, daRocha:2008we, Villalobos:2015xca, Cavalcanti:2014uta,daRocha:2013qhu}, while a spinor such that either $\Theta\!\neq\!0$ or $\Phi\!\neq\!0$ is called \emph{regular} and it is the center of attention of the present work: we can always write the most general regular spinor in terms of a generic complex Lorentz transformation $\boldsymbol{S}$ according to
\begin{eqnarray}
&\!\psi\!=\!\phi e^{-\frac{i}{2}\beta\boldsymbol{\pi}}
\boldsymbol{S}\left(\!\begin{tabular}{c}
$1$\\
$0$\\
$1$\\
$0$
\end{tabular}\!\right)
\label{spinor}
\end{eqnarray}
called \emph{polar form} \cite{Fabbri:2016msm}, and such that with it we have
\begin{eqnarray}
&\!\!M_{ab}\!=\!2i\overline{\psi}\boldsymbol{\sigma}_{ab}\psi
\!=\!2\phi^{2}(\cos{\beta}u^{j}s^{k}\varepsilon_{jkab}\!+\!\sin{\beta}u_{[a}s_{b]})
\end{eqnarray}
in terms of
\begin{eqnarray}
&\!S^{a}\!=\!\overline{\psi}\boldsymbol{\gamma}^{a}\boldsymbol{\pi}\psi\!=\!2\phi^{2}s^{a}\\
&\!U^{a}\!=\!\overline{\psi}\boldsymbol{\gamma}^{a}\psi\!=\!2\phi^{2}u^{a}
\end{eqnarray}
such that $u_{a}u^{a}\!=\!-s_{a}s^{a}\!=\!1$ and $u_{a}s^{a}\!=\!0$ and representing the velocity vector and the spin axial-vector as well as
\begin{eqnarray}
&\Theta\!=\!i\overline{\psi}\boldsymbol{\pi}\psi\!=\!2\phi^{2}\sin{\beta}\\
&\Phi\!=\!\overline{\psi}\psi\!=\!2\phi^{2}\cos{\beta}
\end{eqnarray}
being a scalar and a pseudo-scalar known as module and Yvon-Takabayashi angle and which are the only two real degrees of freedom of the spinor field. Because generally
\begin{eqnarray}
&\boldsymbol{S}\partial_{\mu}\boldsymbol{S}^{-1}\!=\!i\partial_{\mu}\theta\mathbb{I}
\!+\!\frac{1}{2}\partial_{\mu}\theta_{ij}\boldsymbol{\sigma}^{ij}
\end{eqnarray}
where $\theta$ is a generic complex phase and $\theta_{ij}\!=\!-\theta_{ji}$ are the six parameters of the Lorentz group, then we can define
\begin{eqnarray}
&\partial_{\mu}\theta_{ij}\!-\!\Omega_{ij\mu}\!\equiv\!R_{ij\mu}\label{R}\\
&\partial_{\mu}\theta\!-\!qA_{\mu}\!\equiv\!P_{\mu}\label{P}
\end{eqnarray}
being real tensors called \emph{tensorial connection} and \emph{gauge vector momentum} respectively, with which we have
\begin{eqnarray}
&\!\boldsymbol{\nabla}_{\mu}\psi\!=\!(\nabla_{\mu}\ln{\phi}\mathbb{I}
\!-\!\frac{i}{2}\nabla_{\mu}\beta\boldsymbol{\pi}
\!-\!iP_{\mu}\mathbb{I}\!-\!\frac{1}{2}R_{ij\mu}\boldsymbol{\sigma}^{ij})\psi
\label{decspinder}
\end{eqnarray}
and also
\begin{eqnarray}
&\nabla_{\mu}s_{i}\!=\!R_{ji\mu}s^{j}\label{ds}\\
&\nabla_{\mu}u_{i}\!=\!R_{ji\mu}u^{j}\label{du}
\end{eqnarray}
identically: from the commutator of the covariant derivatives we deduce that the curvatures are such that
\begin{eqnarray}
&\!\!\!\!\!\!\!\!R^{i}_{\phantom{i}j\mu\nu}\!=\!-(\nabla_{\mu}R^{i}_{\phantom{i}j\nu}
\!-\!\!\nabla_{\nu}R^{i}_{\phantom{i}j\mu}
\!\!+\!R^{i}_{\phantom{i}k\mu}R^{k}_{\phantom{k}j\nu}
\!-\!R^{i}_{\phantom{i}k\nu}R^{k}_{\phantom{k}j\mu})\label{Riemann}\\
\!\!\!\!&qF_{\mu\nu}\!=\!-(\nabla_{\mu}P_{\nu}\!-\!\nabla_{\nu}P_{\mu})\label{Maxwell}
\end{eqnarray}
telling that the tensors defined in (\ref{R}, \ref{P}) do not generate any curvature tensor that is not already generated by the spin connection and the gauge potential. The Lagrangian above gives rise to field equations that in the polar form are transcribed into the geometric field equations
\begin{eqnarray}
\nonumber
&\nabla_{k}R^{ka}_{\phantom{ka}a}g^{\rho\sigma}\!-\!\nabla_{i}R^{i\sigma\rho}
\!-\!\nabla^{\rho}R^{\sigma i}_{\phantom{\sigma i}i}\!+\!R_{ki}^{\phantom{ki}i}R^{k\sigma\rho}+\\
\nonumber
&+R_{ik}^{\phantom{ik}\rho}R^{k\sigma i}
\!-\!\frac{1}{2}R_{ki}^{\phantom{ki}i}R^{ka}_{\phantom{ka}a}g^{\rho\sigma}-\\
\nonumber
&-\frac{1}{2}R^{ika}R_{kai}g^{\rho\sigma}\!=\!\frac{1}{2}[M^{2}(W^{\rho}W^{\sigma}
\!\!-\!\!\frac{1}{2}W^{\alpha}W_{\alpha}g^{\rho\sigma})+\\
\nonumber
&+\frac{1}{4}(\partial W)^{2}g^{\rho\sigma}
\!-\!(\partial W)^{\sigma\alpha}(\partial W)^{\rho}_{\phantom{\rho}\alpha}+\\
\nonumber
&+\frac{1}{4}F^{2}g^{\rho\sigma}\!-\!F^{\rho\alpha}\!F^{\sigma}_{\phantom{\sigma}\alpha}-\\
\nonumber
&-\phi^{2}[(XW\!-\!\nabla\frac{\beta}{2})^{\sigma}s^{\rho}
\!+\!(XW\!-\!\nabla\frac{\beta}{2})^{\rho}s^{\sigma}-\\
\nonumber
&-P^{\sigma}u^{\rho}\!-\!P^{\rho}u^{\sigma}+\\
&+\frac{1}{4}R_{ij}^{\phantom{ij}\sigma}\varepsilon^{\rho ijk}s_{k}
\!+\!\frac{1}{4}R_{ij}^{\phantom{ij}\rho}\varepsilon^{\sigma ijk}s_{k}]]\label{ee}
\end{eqnarray}
with
\begin{eqnarray}
&\nabla^{2}P^{\mu}
\!-\!\nabla_{\sigma}\nabla^{\mu}P^{\sigma}\!=\!-2q^{2}\phi^{2}u^{\mu}\label{me}
\end{eqnarray}
and
\begin{eqnarray}
&\!\!\!\!\nabla^{2}(XW)^{\mu}\!-\!\nabla_{\alpha}\nabla^{\mu}(XW)^{\alpha}
\!+\!M^{2}XW^{\mu}\!=\!2X^{2}\phi^{2}s^{\mu}\label{se}
\end{eqnarray}
alongside to the matter field equations
\begin{eqnarray}
\nonumber
&\frac{1}{2}\varepsilon_{\mu\alpha\nu\iota}R^{\alpha\nu\iota}
\!-\!2P^{\iota}u_{[\iota}s_{\mu]}+\\
&+2(\nabla\beta/2\!-\!XW)_{\mu}\!+\!2s_{\mu}m\cos{\beta}\!=\!0\label{dep1}\\
\nonumber
&R_{\mu a}^{\phantom{\mu a}a}
\!-\!2P^{\rho}u^{\nu}s^{\alpha}\varepsilon_{\mu\rho\nu\alpha}+\\
&+2s_{\mu}m\sin{\beta}\!+\!\nabla_{\mu}\ln{\phi^{2}}\!=\!0\label{dep2}
\end{eqnarray}
specifying all the first-order derivatives of the module and the YT angle \cite{h1}, and which can be proven to be equivalent to the original Dirac spinor field equations \cite{Fabbri:2016laz}.

To see that, we start by considering the Lagrangian we have written in (\ref{l}) and then we vary it with respect to the spinor field, getting the field equations
\begin{eqnarray}
&i\boldsymbol{\gamma}^{\mu}\boldsymbol{\nabla}_{\mu}\psi
\!-\!XW_{\mu}\boldsymbol{\gamma}^{\mu}\boldsymbol{\pi}\psi\!-\!m\psi\!=\!0\label{D}
\end{eqnarray}
which we then multiply by $\boldsymbol{\gamma}^{a}\boldsymbol{\pi}$ and $\boldsymbol{\gamma}^{a}$ and by the conjugate spinor, splitting real and imaginary parts, to get the four real vectorial field equations given according to
\begin{eqnarray}
\nonumber
&i(\overline{\psi}\boldsymbol{\nabla}^{\alpha}\psi
\!-\!\boldsymbol{\nabla}^{\alpha}\overline{\psi}\psi)
\!-\!\nabla_{\mu}M^{\mu\alpha}-\\
&-XW_{\sigma}M_{\mu\nu}\varepsilon^{\mu\nu\sigma\alpha}\!-\!2mU^{\alpha}\!=\!0
\label{vr}\\
\nonumber
&\nabla_{\alpha}\Phi
\!-\!2(\overline{\psi}\boldsymbol{\sigma}_{\mu\alpha}\!\boldsymbol{\nabla}^{\mu}\psi
\!-\!\!\boldsymbol{\nabla}^{\mu}\overline{\psi}\boldsymbol{\sigma}_{\mu\alpha}\psi)+\\
&+2X\Theta W_{\alpha}\!=\!0\label{vi}\\
\nonumber
&\nabla_{\nu}\Theta\!-\!
2i(\overline{\psi}\boldsymbol{\sigma}_{\mu\nu}\boldsymbol{\pi}\boldsymbol{\nabla}^{\mu}\psi\!-\!
\boldsymbol{\nabla}^{\mu}\overline{\psi}\boldsymbol{\sigma}_{\mu\nu}\boldsymbol{\pi}\psi)-\\
&-2X\Phi W_{\nu}\!+\!2mS_{\nu}\!=\!0\label{ar}\\
\nonumber
&(\boldsymbol{\nabla}_{\alpha}\overline{\psi}\boldsymbol{\pi}\psi
\!-\!\overline{\psi}\boldsymbol{\pi}\boldsymbol{\nabla}_{\alpha}\psi)
\!-\!\frac{1}{2}\nabla^{\mu}M^{\rho\sigma}\varepsilon_{\rho\sigma\mu\alpha}+\\
&+2XW^{\mu}M_{\mu\alpha}\!=\!0\label{ai}
\end{eqnarray}
and in which we now plug the polar form (\ref{spinor}) obtaining
\begin{eqnarray}
\nonumber
&-\nabla_{\mu}\ln{\phi}M^{\mu\sigma}
\!+\!\frac{1}{2}(\frac{1}{2}\nabla_{\mu}\beta\!-\!XW_{\mu})M_{\pi\nu}
\varepsilon^{\pi\nu\mu\sigma}+\\
\nonumber
&+P^{\sigma}\Phi\!+\!\frac{1}{8}R^{\alpha\nu\rho}M_{\pi\kappa}
\varepsilon_{\alpha\nu\rho\mu}\varepsilon^{\pi\kappa\sigma\mu}-\\
&-\frac{1}{2}R_{\mu a}^{\phantom{\mu a}a}M^{\mu\sigma}\!-\!mU^{\sigma}\!=\!0\\
\nonumber
&-\nabla_{\sigma}\ln{\phi}\Phi\!+\!(\frac{1}{2}\nabla_{\sigma}\beta\!-\!XW_{\sigma})\Theta
\!-\!P^{\mu}M_{\mu\sigma}-\\
&-\frac{1}{4}R^{\alpha\nu\rho}\varepsilon_{\alpha\nu\rho\sigma}\Theta
\!-\!\frac{1}{2}R_{\sigma a}^{\phantom{\sigma a}a}\Phi\!=\!0\\
\nonumber
&\nabla_{\sigma}\ln{\phi}\Theta\!+\!(\frac{1}{2}\nabla_{\sigma}\beta\!-\!XW_{\sigma})\Phi+\\
\nonumber
&+\frac{1}{2}P^{\mu}M^{\pi\kappa}\varepsilon_{\pi\kappa\mu\sigma}
\!-\!\frac{1}{4}R^{\alpha\nu\rho}\varepsilon_{\alpha\nu\rho\sigma}\Phi+\\
&+\frac{1}{2}R_{\sigma a}^{\phantom{\sigma a}a}\Theta\!+\!mS_{\sigma}\!=\!0\\
\nonumber
&\frac{1}{2}\nabla_{\mu}\ln{\phi}M_{\pi\kappa}\varepsilon^{\pi\kappa\mu\sigma}
\!+\!(\frac{1}{2}\nabla_{\mu}\beta\!-\!XW_{\mu})M^{\mu\sigma}-\\
&-P^{\sigma}\Theta\!-\!\frac{1}{4}R^{\alpha\nu\rho}M^{\mu\sigma}\varepsilon_{\alpha\nu\rho\mu}
\!+\!\frac{1}{4}R_{\mu a}^{\phantom{\mu a}a}M_{\pi\kappa}\varepsilon^{\pi\kappa\mu\sigma}\!=\!0
\end{eqnarray}
as a straightforward substitution shows: the second and third, after inserting the bi-linear quantities, become
\begin{eqnarray}
\nonumber
&\frac{1}{2}\nabla_{\alpha}\ln{\phi^{2}}\cos{\beta}
\!-\!(\frac{1}{2}\nabla_{\alpha}\beta\!-\!XW_{\alpha})\sin{\beta}+\\
\nonumber
&+P^{\mu}(u^{\rho}s^{\sigma}\varepsilon_{\rho\sigma\mu\alpha}\cos{\beta}
\!+\!u_{[\mu}s_{\alpha]}\sin{\beta})+\\
&+\frac{1}{2}R_{\alpha\mu}^{\phantom{\alpha\mu}\mu}\cos{\beta}
\!+\!\frac{1}{4}R^{\rho\sigma\mu}\varepsilon_{\rho\sigma\mu\alpha}\sin{\beta}\!=\!0\\
\nonumber
&\frac{1}{2}\nabla_{\nu}\ln{\phi^{2}}\sin{\beta}
\!+\!(\frac{1}{2}\nabla_{\nu}\beta\!-\!XW_{\nu})\cos{\beta}+\\
\nonumber
&+P^{\mu}(u^{\rho}s^{\sigma}\varepsilon_{\rho\sigma\mu\nu}\sin{\beta}\!-\!u_{[\mu}s_{\nu]}\cos{\beta})-\\
&-\frac{1}{4}R^{\rho\sigma\mu}\varepsilon_{\rho\sigma\mu\nu}\cos{\beta}
\!+\!\frac{1}{2}R_{\nu\mu}^{\phantom{\nu\mu}\mu}\sin{\beta}\!+\!ms_{\nu}\!=\!0
\end{eqnarray}
and after diagonalization
\begin{eqnarray}
\nonumber
&\frac{1}{2}\varepsilon_{\mu\alpha\nu\iota}R^{\alpha\nu\iota}
\!-\!2P^{\iota}u_{[\iota}s_{\mu]}-\\
&-2XW_{\mu}\!+\!\nabla_{\mu}\beta\!+\!2s_{\mu}m\cos{\beta}\!=\!0\\
\nonumber
&R_{\mu a}^{\phantom{\mu a}a}
\!-\!2P^{\rho}u^{\nu}s^{\alpha}\varepsilon_{\mu\rho\nu\alpha}+\\
&+2s_{\mu}m\sin{\beta}\!+\!\nabla_{\mu}\ln{\phi^{2}}\!=\!0
\end{eqnarray}
in general. Conversely, from these and then considering the general identities given by the expressions 
\begin{eqnarray}
&2\boldsymbol{\sigma}^{\mu\nu}u_{\mu}s_{\nu}\boldsymbol{\pi}\psi\!+\!\psi=0\\
&is_{\mu}\boldsymbol{\gamma}^{\mu}\psi\sin{\beta}
\!+\!s_{\mu}\boldsymbol{\gamma}^{\mu}\boldsymbol{\pi}\psi\cos{\beta}\!+\!\psi=0
\end{eqnarray}
it is possible to see that
\begin{eqnarray}
\nonumber
&i\boldsymbol{\gamma}^{\mu}\boldsymbol{\nabla}_{\mu}\psi
\!-\!XW_{\sigma}\boldsymbol{\gamma}^{\sigma}\boldsymbol{\pi}\psi\!-\!m\psi=\\
\nonumber
&=[i\boldsymbol{\gamma}^{\mu}P^{\rho}u^{\nu}s^{\alpha}\varepsilon_{\mu\rho\nu\alpha}+\\
\nonumber
&+P^{\iota}u_{[\iota}s_{\mu]}\boldsymbol{\gamma}^{\mu}\boldsymbol{\pi}
\!+\!P_{\mu}\boldsymbol{\gamma}^{\mu}-\\
&-is_{\mu}\boldsymbol{\gamma}^{\mu}m\sin{\beta}
\!-\!s_{\mu}\boldsymbol{\gamma}^{\mu}\boldsymbol{\pi}m\cos{\beta}\!-\!m\mathbb{I}]\psi\!=\!0
\end{eqnarray}
showing that when the spinor is in polar form these field equations are valid: as any spinor can always be written in polar form then also these field equations are valid in general. So (\ref{D}) is equivalent to (\ref{dep1}, \ref{dep2}) in general \cite{Fabbri:2017pwp}.

\subsection{Gauge Momentum}
As we already said, the objects (\ref{R}, \ref{P}) are real tensors and, because of the information content that can be deduced from their definition, they contain all information normally contained within the connection and the gauge potential, and thus we called them tensorial connection and gauge vector momentum, respectively: in particular we have that $R_{ijk}$ has a trace defined as
\begin{eqnarray}
&\!R_{a}\!=\!R_{ac}^{\phantom{ac}c}
\end{eqnarray}
and its completely antisymmetric part has dual as
\begin{eqnarray}
&\!\!B_{a}\!=\!\frac{1}{2}\varepsilon_{aijk}R^{ijk}
\end{eqnarray}
so that the non-completely antisymmetric traceless part
\begin{eqnarray}
&\!\!\!\!\Pi_{ijk}\!=\!R_{ijk}\!-\!\frac{1}{3}(R_{i}\eta_{jk}\!-\!R_{j}\eta_{ik})
\!-\!\frac{1}{3}\varepsilon_{ijka}B^{a}
\end{eqnarray}
is such that $\Pi_{ia}^{\phantom{ia}a}\!=0$ and $\Pi_{ijk}\varepsilon^{ijka}\!=0$ hold; instead, for $P_{a}$ we have irreducibility. However, it is possible to have $P_{a}$ written in terms of $R_{ijk}$ according to the expression
\begin{eqnarray}
\nonumber
&P^{\mu}\!=\!m\cos{\beta}u^{\mu}
\!-\!\frac{1}{2}(\nabla_{k}\beta\!-\!2XW_{k}\!+\!B_{k})s^{[k}u^{\mu]}-\\
&-\frac{1}{2}(\nabla_{k}\ln{\phi^{2}}\!+\!R_{k})s_{j}u_{i}\varepsilon^{kji\mu}\label{momentum}
\end{eqnarray}
although it is only a link between $P_{a}$ and the two vectorial parts of $R_{ijk}$ and not a link to the full tensor; this is clear, because the only occurrence of the full $R_{ijk}$ is within the field equations (\ref{ee}) but even there it is always either in derivatives or in products. Only 
(\ref{dep1}, \ref{dep2}) contain the pure forms of $R_{a}$ and $B_{a}$ needed for (\ref{momentum}) to be expressed.

To see that, consider (\ref{dep1}, \ref{dep2}) in terms of $R_{a}$ and $B_{a}$
\begin{eqnarray}
&\!\!B_{\mu}\!-\!2P^{\iota}u_{[\iota}s_{\mu]}\!+\!2(\nabla\beta/2\!-\!XW)_{\mu}
\!+\!2s_{\mu}m\cos{\beta}\!=\!0\\
&\!R_{\mu}\!-\!2P^{\rho}u^{\nu}s^{\alpha}\varepsilon_{\mu\rho\nu\alpha}
\!+\!2s_{\mu}m\sin{\beta}\!+\!\nabla_{\mu}\ln{\phi^{2}}\!=\!0
\end{eqnarray}
and then contract the first by $u^{\mu}$ and $s^{\mu}$ and the second by $u^{\nu}s^{\alpha}\varepsilon_{\nu\alpha\mu\rho}$ eventually getting 
\begin{eqnarray}
&Ps\!+\!\frac{1}{2}(\nabla\beta\!-\!2XW\!+\!B)u\!=\!0\\
&Pu\!+\!\frac{1}{2}(\nabla\beta\!-\!2XW\!+\!B)s\!-\!m\cos{\beta}\!=\!0\\
&\!P^{\rho}\!+\!Ps s^{\rho}\!-\!Pu u^{\rho}
\!+\!\frac{1}{2}(\nabla\ln{\phi^{2}}\!+\!R)_{\mu}s_{\alpha}u_{\nu}
\varepsilon^{\mu\alpha\nu\rho}\!=\!0
\end{eqnarray}
which are now easier to manipulate since in the last one $P^{\rho}$ appears isolated and the other occurrences $Ps$ and $Pu$ can be substituted in terms of the other two expressions given above: if the replacement is made then we obtain
\begin{eqnarray}
\nonumber
&P^{\rho}\!=\!\frac{1}{2}(\nabla\beta\!-\!2XW\!+\!B)u s^{\rho}
\!-\!\frac{1}{2}(\nabla\beta\!-\!2XW\!+\!B)s u^{\rho}+\\
&+m\cos{\beta}u^{\rho}\!-\!\frac{1}{2}(\nabla\ln{\phi^{2}}\!+\!R)_{\mu}s_{\alpha}u_{\nu}
\varepsilon^{\mu\alpha\nu\rho}
\end{eqnarray}
in general. Therefore (\ref{dep1}, \ref{dep2}) imply (\ref{momentum}) in general \cite{Fabbri:2018crr}.

\subsection{Velocity}
Expression (\ref{momentum}) gives the gauge momentum in terms of the module and the Yvon-Takabayashi angle, but also in terms of the velocity and the spin: because both momentum and spin are supposed to be constants of motion, it is interesting to invert it for the velocity. For this, define
\begin{eqnarray}
&m\cos{\beta}\!-\!\frac{1}{2}(\nabla\beta\!-\!2XW\!+\!B)_{k}s^{k}\!=\!X\\
&\frac{1}{2}(\nabla\beta\!-\!2XW\!+\!B)_{k}\!=\!Y_{k}\\
&-\frac{1}{2}(\nabla\ln{\phi^{2}}\!+\!R)_{k}\!=\!Z_{k}
\end{eqnarray}
in terms of which the momentum is written as
\begin{eqnarray}
&P^{a}\!=\!(X\eta^{ak}\!+\!Y^{k}s^{a}\!+\!Z_{i}s_{j}\varepsilon^{ijka})u_{k}\label{a}
\end{eqnarray}
in the form of a matrix containing only the spin applied to the velocity: when inverted it will give the velocity as a product of a specific spin-dependent matrix applied to the momentum. To get such inversion, have (\ref{a}) dotted into $Z^{i}s^{j}\varepsilon_{ijka}$ and $Z_{a}$ so to obtain
\begin{eqnarray}
\nonumber
&P^{a}Z^{i}s^{j}\varepsilon_{ijka}\!=\!XZ^{i}s^{j}u^{a}\varepsilon_{ijka}+\\
&+(Z^{2}\!+\!|Z\!\cdot\!s|^{2})u_{k}\!-\!Z\!\cdot\!u (Z_{k}\!+\!Z\!\cdot\!s s_{k})\label{b1}\\
&P\!\cdot\!Z\!+\!P\!\cdot\!s Z\!\cdot\!s\!=\!XZ\!\cdot\!u\label{b2}
\end{eqnarray}
and after having (\ref{b2}) substituted into (\ref{b1}) and the result substituted back into (\ref{a}) we finally end up with
\begin{eqnarray}
\nonumber
&u^{k}\!=\!(1\!+\!Z^{2}/X^{2}\!+\!|Z\!\cdot\!s|^{2}/X^{2})^{-1}[\eta^{ka}+\\
\nonumber
&+s^{a}s^{k}(1\!+\!|Z\!\cdot\!s|^{2}/X^{2})\!+\!Z^{a}Z^{k}/X^{2}+\\
&+(s^{a}Z^{k}\!+\!Z^{a}s^{k})Z\!\cdot\!s/X^{2}\!+\!Z_{i}s_{j}\varepsilon^{ijka}/X]P_{a}/X
\end{eqnarray}
giving the velocity as product of a specific spin-dependent matrix further applied onto the momentum in general.

More specifically we can introduce also
\begin{eqnarray}
&\zeta_{k}\!=\!Z_{k}/X
\end{eqnarray}
and write
\begin{eqnarray}
\nonumber
&u^{k}\!=\!(1\!+\!\zeta^{2}\!+\!|\zeta\!\cdot\!s|^{2})^{-1}[\eta^{ka}+\\
\nonumber
&+s^{a}s^{k}(1\!+\!|\zeta\!\cdot\!s|^{2})\!+\!\zeta^{a}\zeta^{k}+\\
&+(s^{a}\zeta^{k}\!+\!\zeta^{a}s^{k})\zeta\!\cdot\!s\!+\!\zeta_{i}s_{j}\varepsilon^{ijka}]P_{a}/X
\label{eq}
\end{eqnarray}
as the most compact form we can have again in general.
\section{Physical Effects}

\subsection{Internal Dynamics}
In the previous section, we have introduced the polar form of spinors in $4$ dimensions. Of course, one may also consider what the polar form would look like when time is a parameter, and so in $3$ dimensions: in such a case, a spinor would have two complex components, hence $4$ real components, and because the spinor transformation law would contain only $3$ rotations, up to $3$ of these components can be removed. This spinor in polar form would have a single degree of freedom, the module. Therefore, if the non-relativistic spinor ought be obtained as a limit of the general spinor, this limit must account for the fact that the Yvon-Takabayashi angle has to vanish beside the fact that the velocity spatial part has to vanish \cite{Fabbri:2016msm,Fabbri:2016laz}.

Because when the spatial part of the velocity is equal to zero but the Yvon-Takabayashi angle is different from zero we are still unable to obtain the full non-relativistic limit, then we must conclude that the Yvon-Takabayashi angle describes the motion of what remains even in rest frame, which has to be the intrinsic motion. That is, it is the motion describing the internal dynamics of a spinor.

We notice that in the definition of the non-relativistic limit, given by the $\beta\!\rightarrow\!0$ and $\vec{u}\!\rightarrow\!0$ above, there appears no gauge momentum (\ref{momentum}) at all: as a matter of fact, the momentum in non-relativistic limit is given by
\begin{eqnarray}
&E\!=\!m\!-\!(X\vec{W}\!-\!\frac{1}{2}\vec{B})\!\cdot\!\vec{s}\\
&\vec{P}\!=\!-(XW^{0}\!-\!\frac{1}{2}B^{0})\vec{s}
\!-\!\frac{1}{2}(\vec{\nabla}\ln{\phi^{2}}\!-\!\vec{R})\!\times\!\vec{s}
\end{eqnarray}
showing that the energy does not reduce to the mass and the spatial momentum does not vanish. The usual limit given by $P_{a}\!\rightarrow\!(m,\vec{0})$ can only be obtained by neglecting the spin content of the spinor, that is if the macroscopic approximation is also implemented for field distributions.

Again, this makes sense, because non-relativistic limit means small spatial momentum only if the momentum is free from spin contributions, and that is from the internal dynamics. Therefore, it is reasonable that such limit can be obtained in this way only by requiring that the internal dynamics be concealed inside the field distribution, as it would normally happen for macroscopic approximations.

Having obtained some insight from the non-relativistic limit, let us see what happens for velocities that are large in general. In this case we can use the expression of the velocity given by (\ref{eq}) in terms of gauge momentum and spin while depending on module and Yvon-Takabayashi angle in general. Considering that momentum and spin are constants of motion, module and Yvon-Takabayashi angle are the only variables. This tells us that spinorial fields can be seen as very peculiar types of fluid for which the velocity depends on density and internal dynamics.

Nonetheless, (\ref{eq}) is too complicated to get meaningful information. To simplify, we study limiting cases, and as special case we take $\zeta_{a}$ to be small so that we have
\begin{eqnarray}
&u^{k}\!\approx\!(\eta^{ka}\!+\!s^{a}s^{k}\!+\!\zeta_{i}s_{j}\varepsilon^{ijka})P_{a}/X
\end{eqnarray}
to the first-order perturbative. Notice that the $\zeta_{a}$ potential contains the gradient of the logarithm of the density, so it can be regarded as the de Broglie-Bohm quantum potential, in the first-order derivative form, and containing also the Yvon-Takabayashi angle contributions.

Differently from the non-relativistic case, based on the Schr\"{o}dinger equation, for which the quantum potential is second-order derivative, in this most general case, based on Dirac equations, the quantum potential is first-order derivative. And differently from the non-relativistic case, where only the module is present, in this most general of cases, both the module and the Yvon-Takabayashi angle give some contribution to the quantum potential.

As a consequence, we regard the $\zeta_{a}$ vector as the quantum potential in the most general form possible.

If we were to take the approximation of small $Y_{a}$ then
\begin{eqnarray}
&\!\!m\cos{\beta}u^{k}\!\approx\!\left(1\!+\!\frac{Y_{b}s^{b}}{m\cos{\beta}}\right)\!
(\eta^{ka}\!+\!s^{a}s^{k}\!+\!\zeta_{i}s_{j}\varepsilon^{ijka})P_{a}
\end{eqnarray}
and for small Yvon-Takabayashi angle
\begin{eqnarray}
&mu^{k}\!\approx\!(1\!+\!Y_{b}s^{b}/m)
(P^{k}\!+\!P_{a}s^{a}s^{k}\!-\!P_{a}\zeta_{i}s_{j}\varepsilon^{aijk})
\end{eqnarray}
with a contribution scaling the momentum in terms of the Yvon-Takabayashi angle plus a contribution changing the direction of the momentum in terms of both module and Yvon-Takabayashi angle. That is, as quantum potential.

Considering only the spatial part, cases of small spatial momentum allow us to better see the effects of the spin contributions. In such cases we have
\begin{eqnarray}
&\vec{u}\!\approx\!(1\!-\!\vec{Y}\!\cdot\!\vec{s}/m)\vec{Z}\!\times\!\vec{s}/m
\end{eqnarray}
with the spin contributions that are in fact explicit.

If torsion were negligible and $B_{a}\!=\!R_{a}\!=\!0$ \cite{Fabbri:2017pwp} then
\begin{eqnarray}
&\vec{u}\!\approx\!(1\!+\!\vec{\varsigma}\!\cdot\!\vec{\nabla}\beta/m)
\vec{\nabla}\ln{\phi^{2}}\!\times\!\vec{\varsigma}/m
\end{eqnarray}
having introduced $\vec{\varsigma}\!=\!\vec{s}/2$ as the usual expression of spin.

It is interesting to notice that in this case we can compute the magnetic moment obtaining the expression
\begin{eqnarray}
\nonumber
\vec{\mu}\!=\!\frac{1}{2}\!\int\!\vec{r}\!\times\!q\vec{U}dV=\\
\nonumber
=\frac{q}{2m}2\!\int\!(1\!+\!\vec{\varsigma}\!\cdot\!\vec{\nabla}\beta/m)
[\vec{r}\!\times\!(\vec{\nabla}\phi^{2}\!\times\!\vec{\varsigma}\,)]dV=\\
\nonumber
=\frac{q}{2m}2\!\int\!(1\!+\!\vec{\varsigma}\!\cdot\!\vec{\nabla}\beta/m)
2\phi^{2}\vec{\varsigma}dV=\\
=\frac{q}{2m}2\vec{\varsigma}\left(1\!+\!\langle\vec{\varsigma}\!\cdot\!\vec{\nabla}\beta/m\rangle\right)
\end{eqnarray}
and so that we can finally write
\begin{eqnarray}
&(g\!-\!2)/2\!\approx\!\langle\vec{\varsigma}\!\cdot\!\vec{\nabla}\beta/m\rangle
\label{correction}
\end{eqnarray}
in terms of the common form of the gyromagnetic factor.

To first order, this would agree with the $\alpha/2\pi$ term if on average the gradient of $\beta$ along the spin were equal to the fine-structure constant $\alpha$ over half Compton length, and this is remarkable since $\beta$ is generally expected to be of the order of the fine-structure constant $\alpha$ \cite{Fabbri:2018crr}.

On the other hand, going beyond order-of-magnitude evaluations requires the Yvon-Takabayashi angle $\beta$ to be known in terms of exact solutions and for the time being this task appears to be out of the possibilities.

More in general, the spin contributions are as in
\begin{eqnarray}
&\vec{u}\!\approx\!(1\!-\!\vec{Y}\!\cdot\!\vec{s}/m)\vec{Z}\!\times\!\vec{s}/m
\end{eqnarray}
showing that the correction to the magnetic moment will depend not only on the Yvon-Takabayashi angle but also on torsion and on the $B_{a}$ axial vector, altogether collected into the $Y_{a}$ axial vector. While the very presence of this term depends on the $\zeta_{a}$ vector, it is necessary for further corrections that the terms $X$ or $Y_{a}$ be present as well.

While the $\zeta_{a}$ vector is the quantum potential providing quantum mechanical effects, the $X$ or $Y_{a}$ terms are what provides quantum field theoretical corrections.

\subsection{Chirality}
When we consider again the non-relativistic limit, the first condition is that of boosting into the rest frame, and in this frame the assumption of rotating the spinor so to align its spin along the third axis is equivalent to require that in the polar form $\boldsymbol{S}$ be the identity: in this instance
\begin{eqnarray}
&\!\psi\!=\!\phi\left(\!\begin{tabular}{c}
$e^{\frac{i}{2}\beta}$\\
$0$\\
$e^{-\frac{i}{2}\beta}$\\
$0$
\end{tabular}\!\right)
\end{eqnarray}
in chiral representation or
\begin{eqnarray}
&\!\psi\!=\!\phi \sqrt{2}
\!\left(\!\begin{tabular}{c}
$\cos{\frac{\beta}{2}}$\\
$0$\\
$-i\sin{\frac{\beta}{2}}$\\
$0$
\end{tabular}\!\right)
\end{eqnarray}
in standard representation. The remaining condition demands that the Yvon-Takabayashi angle vanishes
\begin{eqnarray}
\!\psi\!=\!\phi \sqrt{2}
\!\left(\!\begin{tabular}{c}
$1$\\
$0$\\
$0$\\
$0$
\end{tabular}\!\right)
\end{eqnarray}
showing that the lower component is zero. This allows a single condition to represent the non-relativistic limit.

When the spinor is written in standard representation, its lower component is for this reason also called small component. As for the same spinor in chiral representation, the lower component is the right-handed component and the upper component is the left-handed component.

So the Yvon-Takabayashi angle, which is related to the small component, gives the phase opposition between the right-handed and left-handed components of spinors.

Because zitterbewegung effects are known to arise from the existence of the small component, or more in general from the interplay of the chiral components, then we can infer that zitterbewegung must be linked to the presence of the Yvon-Takabayashi angle quite generally.

It is also interesting to see that the bi-linear quantities $\Theta$ and $\Phi$ when written in terms of the left-handed and the right-handed components $L$ and $R$ have expressions given by $\Theta\!=\!i(L^{\dagger}R\!-\!R^{\dagger}L)$ and $\Phi\!=\!(L^{\dagger}R\!+\!R^{\dagger}L)$ so that, for such a reason, they assume a remarkably clear interpretation.

Considering that the Yvon-Takabayashi angle is related to the scalar $\Theta/\Phi$ while the module is $\Theta^{2}\!+\Phi^{2}$ we interpret the module and the Yvon-Takabayashi angle as the mean of the chiral components and the standard deviation from the mean of the chiral components, respectively.

This recovers the interpretation of the module and the Yvon-Takabayashi angle as describing an averaged material distribution with its internal structure.
\section{Applications}

\subsection{Anomaly}
Having the expression of the velocity (\ref{eq}), an application of some importance can be found for the electrodynamic potential. This potential has the form
\begin{eqnarray}
&\mathscr{V}\!=\!qU^{k}A_{k}
\end{eqnarray}
and consequently we can write
\begin{eqnarray}
\nonumber
&\mathscr{V}\!=\!2q\phi^{2}(1\!+\!\zeta^{2}\!+\!|\zeta\!\cdot\!s|^{2})^{-1}[P_{a}A^{a}+\\
\nonumber
&+P_{a}s^{a}A_{k}s^{k}(1\!+\!|\zeta\!\cdot\!s|^{2})\!+\!P_{a}\zeta^{a}A_{k}\zeta^{k}+\\
\nonumber
&+(P_{a}s^{a}A_{k}\zeta^{k}\!+\!P_{a}\zeta^{a}A_{k}s^{k})\zeta\!\cdot\!s+\\
&+P_{a}A_{k}\zeta_{i}s_{j}\varepsilon^{ijka}]/X
\label{edp}
\end{eqnarray}
quite straightforwardly. As (\ref{eq}) is the expression of the most general form of the velocity, then it is clear that the above is the most comprehensive electrodynamic potential density. Its integral over the volume must contain the complete information about all electrodynamic effects.

In the previous section we have seen that in some approximation, it is possible to find the magnetic moment correction (\ref{correction}), which should furnish the correct value in specific cases of Yvon-Takabayashi angle. Just the same, we also commented that this can be done only when exact solutions are found, and such task is difficult in general.

Nonetheless, since (\ref{edp}) is the electrodynamic potential developed in its most general form then it should contain full information about all electrodynamic effects.

In the simplest from for plane waves it reduces to
\begin{eqnarray}
&\mathscr{V}_{\mathrm{pw}}\!=\!2q\phi^{2}P_{a}A^{a}/m
\end{eqnarray}
which means that $\Delta\mathscr{V}\!=\!\mathscr{V}\!-\!\mathscr{V}_{\mathrm{pw}}$ contains the information about all effects arising from general solutions and which would not be obtained by analyses employing plane wave solutions. So $\Delta\mathscr{V}$ has the information found in non-zero Yvon-Takabayashi angles or in non-constant modules.

That $\Delta\mathscr{V}$ may encode electrodynamic effects not obtainable with the use of plane waves means that in it we might find all of the anomalous terms due to the radiative corrections normally arising from field quantization.

This idea is not a new suggestion, as anomalies to the gyromagnetic factor have already been studied by using affine structures or spinorial interactions \cite{Capozziello:2014gua,Cirilo-Lombardo:2014opa}.
\section{Conclusion}
In this paper, we have studied the Dirac spinor field in polar form giving the field equations, combining them as to get the explicit form of the momentum, and inverting it as to obtain the explicit form of the velocity: with such tools, we established that the Yvon-Takabayashi angle is what describes the internal dynamics, defined as relative motion between chiral parts, that this is connected to the effects of zitterbewegung, arising from spin contributions, and that both Yvon-Takabayashi angle and module come together to form the $\zeta_{a}$ vector, which is the most general form of quantum potential providing quantum mechanical effects; we also established that the Yvon-Takabayashi angle takes part in the $Y_{a}$ axial vector, which appears to encode the information about the anomalous terms that arise as quantum field theory corrections. Consequently, we have conjectured that the anomalous behaviour fields display might be obtained by considering solutions that are more general than the simplest mere plane waves.

We have shown that this might happen for the anomaly of the magnetic moment, and we have recalled that in the literature it has already been discussed that this may also happen in the case of the running of coupling constants for the renormalization group; other cases were exhibited in the literature mentioned in the introduction. The idea, expressed also in the introduction, that effects described by quantization in terms of plane wave solutions should also be described solely in terms of more general solutions has therefore some evidence, although the actual proof of this conjecture requires exact solutions, which we lack.

Our point in the present paper, however, was much less ambitious. What we wanted to do was merely to present a problem and conjecture a possible way out, that is that the problems of quantization may be altogether circumvented by an approach that does not involve quantization at all but which recovers its effects in terms of more general approaches involving fields displaying more general internal dynamics, and to this purpose, we provided the most general form for the Dirac spinor field theory.

This paper is intended to lay the grounds as a starting point for any future work that aims to face the problem of finding general solutions with internal dynamics which may recover the effects of quantization.

\end{document}